\documentstyle[eqsecnum,aps]{revtex}
\begin{document}
\title{ Thermal Particle and Photon Production \\[.5cm]
in $Pb+Pb$ Collisions with Transverse Flow.}
\author{J. Cleymans, K. Redlich\cite{addKR},    D.K.
Srivastava\cite{Add} }
\address{Department of Physics, University of Cape Town, Rondebosch 7700,
South Africa}
\maketitle
\begin{abstract}
Particle and  photon production  is  analyzed in the
presence of transverse flow using 
two  approximations  to describe the properties of the hadronic
medium,
one containing only  $\pi, \rho, \omega$, and $\eta$
mesons ({\it simplified equation of state}) and the other containing
hadrons and resonances from the particle data table. 
Both are considered with and without initial quark gluon 
plasma formation. In each case the initial 
temperature is fixed by requiring $dN_{ch}/dy \sim$ 550 in the final state.

It is shown that most observables are very sensitive to 
the equation of state.
This is particularly evident 
when comparing the results of the simplified equation of state in the 
scenarios with and without phase transition. The hadronic gas 
scenario leads to a substantially higher rate for the 
$p_T$-distribution
of all particles. 

In the complete equation of state with several hundreds of
hadronic resonances, the difference between the scenarios with and 
without phase transition is rather modest. Both photon and 
particle spectra, in a wide $p_T$ range, show very similar 
behavior. 
It is therefore concluded that from the  
$p_T$ spectra it will be hard to disentangle quark 
gluon plasma formation in the initial state. 
It is to be stressed 
however,  that  there  are  conceptual difficulties in applying
a  pure
hadronic gas equation of state at SPS-energies. 
The phase transition scenario with  a quark 
gluon plasma present in the initial state seems to be the more 
natural one.
\end{abstract}
\section{Introduction}
The   direct  experimental  evidence  for  transverse  flow  in
relativistic  heavy  ion collisions is  weak and in many models it is even
ignored   completely.
The evidence  is mainly ``circumstantial''~\cite{heinz1},
however, in all recent analyses  of
particle  ratios  it  is
impossible  to ignore transverse 
flow without running into contradictions
\cite{nelson,rafelski1,rafelski2,cs,stachel}. There has therefore been
a  lot  of  interest  recently
in  analyzing  the  effects  of  flow more
thoroughly~\cite{rischke1,dks7,rischke2,sollfrank}. 

A   clear illustration of 
the effects of transverse flow is shown in figures 1 and 2.
Figure 1 shows the average transverse momentum as a function of
the  freeze-out  temperature in the absence of flow. 
Clearly if  the freeze-out
temperature  is  very  high  the outgoing particle will be very
energetic  and  the average transverse momentum will be large
accordingly.  However,  in the presence of transverse flow this
is changed completely as 
shown  in  figure 2 where we plot again the transverse momentum
as  a  function  of freeze-out temperature but this time taking
into   account  transverse  flow.  Compared  to  figure  1  the
behaviour  is  changed  totally  :  the  higher  the freeze-out
temperature,  the  lower   the  transverse momentum, just the
opposite from figure 1. The explanation for this
is  very simple. If the freeze-out temperature is very low, the
hadronic  system  will  be long-lived because it takes a longer
time  to  cool down. Since the lifetime is
long,  the  transverse  expansion  will act for a long time, 
 one  ends  up  with  a  large transverse
momentum. The freeze-out
temperature is no longer the deciding factor as it is in figure
1  but  the  lifetime  of  the  hadronic  system  becomes  more
important.  Similarly  the  presence of a phase transition will
also  change  the  transverse  expansion  substantially because
during the phase transition the speed of sound becomes zero and
this  will  slow considerably the transverse expansion. We thus
expect  the  distribution in transverse momentum to be strongly
affected  by  the  lifetime  of the hadronic system and by the
presence (or absence) of a phase transition.

It  is  well   known that the equation of state has an important
effect   on   the   transverse   expansion   of   hot  hadronic
matter~\cite{dks7,sollfrank,dks10}. 
In   particular,   the   occurrence   of   a  phase  transition
considerably  changes the development and the speed with which
the transverse expansion occurs~\cite{rischke1}. 

In  models  of  hydrodynamic  flow  it  is unnatural to have an
instantaneous  freeze-out  of  all particles. This was recently
emphasized  in reference~\cite{rischke2} and follows simply from
the  fact that the freeze-out temperature will affect particles
in  the center of the collision at a later stage than particles
which are at the edges of the hadronic system. It takes some time
for  the  cooling to propagate through the system and thus the
different  freeze-out  times. In  kinetic  models  it is  even
unnatural  for particles of different types to freeze out at the
same time 
because of their different mean free paths. This is illustrated
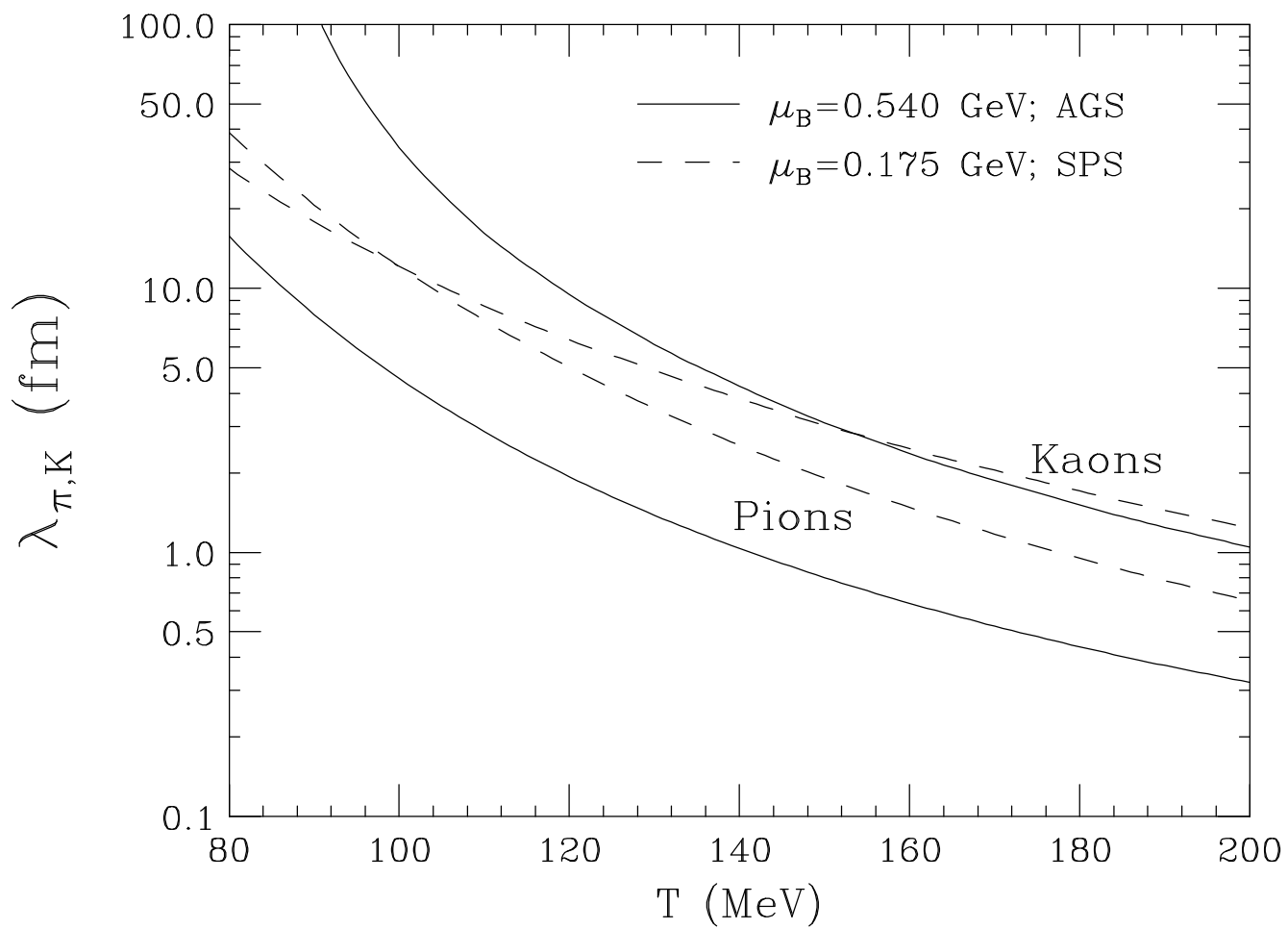
in  figure 3 (based upon reference \cite{haglin}) showing the mean
free paths of pions and kaons as a function of temperature.
In  the  remainder  of  this  paper  we will not take this into
account  and  concentrate  on  the  effects  of  the transverse
expansion  on  particle  production.  We  consider particles to
freeze-out when their environment reaches a certain temperature
and we thus ignore differences between the types of particles.
In this paper we would like to
investigate  in detail the influence of transverse expansion on
as many quantities as possible.
This
in turn affects the transverse momentum distribution. It is the
purpose of this paper to investigate systematically the effect
of  four  different  choices  on the
transverse expansion of hot hadronic matter.

For the hadronic 
matter we will use
two approximations  to describe the medium. 
In the first one  only a small number of mesons 
($\pi, \rho, \omega$, and $\eta$) is used 
({\it simplified equation of state}) while in the second one
all hadrons listed in the  particle data table~\cite{PDT} are used. 
Both equations of state will be considered with, 
and without, a quark gluon 
plasma in the initial state. 
In all four cases  considered the initial 
temperature is fixed by requiring $dN_{ch}/dy \sim$ 550 in the final 
state, assuming $dN/dy =1.5 \times dN_{ch}/dy$
and imposing entropy conservation during the expansion.
The initial thermalization time 
is assumed to be $\tau_{0} \sim$ 1fm.

Most physical observables (but not all) are very sensitive to 
the equation of state, and the assumptions made about the nature of 
the initially created system. This is particularly evident 
when comparing the simplified equation of state in the 
scenarios with and without phase transition. Here the purely hadronic gas 
scenario leads to  substantially larger $p_T$
distributions for all particles (including photons). 
This is mainly due to 
the very different initial temperatures 
arising  from  the  fact  that  the 
freeze-out temperature is the same in all scenarios.

For the equation of state with several hundreds of
hadrons the difference between the scenario with, and
without, phase transition is almost non-existent. 
Thus one concludes that 
it will be  hard to prove the presence of a quark 
gluon plasma in the initial state. 
One has to stress, 
however, 
that the assumption of an
equation of state with the phase transition scenario and  quark 
gluon plasma formation in the initial state is the more 
natural one.
This is because there are conceptual difficulties in applying 
a purely 
hadronic gas equation since a 
temperature $T \sim$ 200 MeV leads to 
a particle density of the order of 2--3/fm$^3$. When considering 
hadrons as extended objects with a  radius $R \sim$ 1~fm, this  
is clearly  not acceptable. 

In  section  2 we briefly review the hydrodynamic equations. In
section  3  we discuss the equation of state in detail. Results
on  the  hadronic  spectra  (pions and kaons) are presented in
section  4.  Photons are discussed separately in section 5. The
$\eta/\pi^0$  ratio  is  discussed  in  section  6.  In section  7
we discuss and summarize our results.
\section{Formulation of Hydrodynamics}
The   standard   starting   point   for  hydrodynamics  is  the
conservation      law      for      the     energy     momentum
tensor~\cite{ruuskanen}
\begin{equation}
\partial^\mu T_{\mu\nu} = 0
\end{equation}
with 
\begin{equation}
T_{\mu\nu}=(\epsilon +P)u_\mu u_\nu -g_{\mu\nu}P
\end{equation}
with the 4-velocity $u_\mu$ given by :
\begin{equation}
u_\mu               =              {1\over\sqrt{1-v^2_r}}
\left(\cosh\eta,v_r\cos\phi,v_r\sin\phi,\sinh\eta\right)  .
\end{equation}
Here $v_r$ is the transverse (radial) velocity, 
$\eta$ is the space-time
rapidity defined as 
\begin{equation}
\eta = {1\over 2}\ln {t+z\over t-z} .
\end{equation}
The  conservation  law $\partial_\mu T^{\mu 0} =0$ leads to the
following equation at zero rapidity :
\begin{eqnarray}
{\partial\over\partial\tau}\left[  {\epsilon +P\over 1-v^2_r} -
P\right] &=& -  {\partial\over\partial  r}\left[ {\epsilon +P\over
1-v^2_r}\right] \nonumber\\
&   &  -  {\epsilon  +P\over  1-v^2_r}  \left[  {v_r\over  r}  +
{1\over\tau}\right]  .
\end{eqnarray}
For numerical purposes is best to rewrite this equation in the form 
\begin{equation}
{\partial\over\partial\tau}T_{00}    +   {\partial\over\partial
r}\left[  v_sT_{00}\right] = -\left[ T_{00} + P\right] \left[
{v_r\over r} + {1\over\tau}\right]
\end{equation}
where
\begin{equation}
v_s = {\epsilon +P\over 1-v^2_r}v_r \Bigg/ \left[ {\epsilon +P\over
1-v^2_r}-P\right]
\end{equation}
This  is  can be readily used in one of the
numerical  algorithms  which have been developed
for solving hydrodynamical equations.

Similarly the equation $\partial_\mu T^{\mu z} =0 $ leads to 
\begin{eqnarray}
{\partial\over\partial\tau}
\left[{\epsilon  +P\over  1-v^2_r}v_r\right] 
&=& -  {\partial\over\partial  r}
\left[  {\epsilon +P\over1-v^2_r}v_r^2\right] \nonumber\\
& & - {\partial P\over\partial r} \nonumber\\
& & -  {\epsilon  +P\over  1-v^2_r} v_r 
\left[{v_r\over  r}  +{1\over\tau}\right] 
\end{eqnarray}
which can be rewritten as 
\begin{equation}
{\partial\over\partial\tau}T_{0r}    +   {\partial\over\partial r}
\left[  v_rT_{0r}\right] = -{\partial P\over\partial r}
 -T_{0r}\left[{v_r\over r} + {1\over\tau}\right] .
\end{equation}
Equations (6) and (10) determine the time evolution of $T_{00}$
and  $T_{0r}$.  From these two we can then determine the energy
density using the following implicit equation 
\begin{equation}
{T_{0r}^2\over T_{00}+P(\epsilon)} = T_{00}-\epsilon .
\end{equation}
The  pressure  is  then determined from the equation of state.
The velocity is determined from $T_{00}$ and $T_{0r}$. The
time  evolution of all the thermodynamic quantities can thus be
determined.
\section{Equation of State}
We discuss in detail four different scenarios. In the first one
we  consider  a  gas  composed  of  a  limited  set  of mesonic
resonances, namely,
$\pi,  \rho, \omega$ and $\eta$ 
(simplified equation of state) while in the second case we use
a  gas  containing  all  resonances listed in the particle data
table\cite{PDT}. Each one of these two cases is then considered
with  or without a phase transition
to a quark-gluon plasma. As has been noted previously
the  presence  of  a  phase transition substantially alters the
transverse   expansion   and  the  distribution  in  transverse
momentum   is   affected   accordingly.  The  phase  transition
slows  down  the  transverse  expansion  and the
average transverse momentum is thus lowered.
The  energy  density  as  a function of temperature is shown in
figure 4. The speed of sound is given in figure 5 as a function
of  the  temperature  while  the  particle  density is shown in
figure  6.  The  pressure  is  given in figure 7. Note that all
these  quantities  are ``raw'' in the sense that no corrections
have  been  made  for  interactions not even in their crudest 
forms like e.g.  excluded volume effects. 

The results in figures 4-7 already show the main differences in
the thermodynamics
of  all  four  equations of state. At the critical temperature,
which  we  assume  to  be  $T_c$ =  160 MeV, the effective
number of degrees of freedom increases by more than a factor of
two  when  going  from  the simplified to the complete particle
input  in  hadronic matter. This naturally reduces the lifetime
of  the  system  in  the  mixed  phase.  A  steeper increase of
pressure  versus  energy  density  seen  in  figure  7  for the
simplified  equation  of  state  implies a higher value for the
speed of sound in figure 5.

The functional dependence of the temperature on proper 
time,  for a purely longitudinal expansion according to Bjorken
hydrodynamics~\cite{bj} seen in
figure 8 shows that with the simplified equation of state with
no  phase  transition one needs a temperature $T\approx $ 320
MeV  to  reproduce  the  final  state entropy. For the complete
equation of state and in the scenario with phase transition the
temperature is reduced to the value of
$\approx$  210  MeV. Thermal particle and photon production is
very  sensitive  to  the temperature value, thus one should see a
very   different   behavior  of  particle  and  photon  spectra
depending  on  the  particle input in the equation of state. In
figure  8 one can also see that the lifetime of the mixed phase
is  much  longer  for  the  simplified  than  for  the complete
equation of state. These results reflecting the effective number
of  degrees  of freedom at $T_c$, imply that for the simplified
version  of the equation of state the thermal photon production
is mostly due to the contribution of the mixed phase. Including
resonances  changes  this  result, reducing the lifetime of the
mixed  phase  and  increasing production from the pure hadronic
phase.  For  small  and  intermediate photon momenta the quark-
gluon plasma contribution to the overall rate is negligible due
to its short lifetime and not to the high initial temperature.

The  freeze-out surface calculated in figures 9-10 shows a very
strong  dependence  on the value of the freeze-out temperature,
which  in the context of hydrodynamical models is considered 
a  free parameter. There are some dynamical restrictions on the
values  of  $T_f$ due to particle scattering in the medium. One
can  see  in figure 3 that for pions a freeze-out temperature
in the range of 100-160 MeV is still allowed as it leads to
a  mean  free  path which is of the order of the  radius
of the $Pb$ nucleus.
Clearly  the  freeze-out  time  is small for particles 
close  to  the  surface  and  largest for the ones close to the
center.  The  higher  initial  temperature leads naturally to a
longer lifetime. For the scenario with
phase   transition   and   simplified  equation  of  state  the
freeze-out  time  is  the  longest  one. This is mostly a
reflection  of  the  long-lived  mixed  phase. All the above
features are contained in figures 9-10.
%
%\section{Freeze-Out and Initial Conditions}
%
%The   initial  conditions  are  as  follows.  
%In  each  one  of the four scenarios considered here, we always
%start from an initial proper time given by 1 fm/c.
%The density of
%charged particles in the
%rapidity distribution is taken as
%\begin{equation}
%{dN\over dy} = 550
%\end{equation}
%%
%The initial temperature is determined by using the conservation
%of entropy:
%The  total  entropy  contained  at  time $\tau_0$ in a slice of
%fluid which occupies a rapidity range $\Delta$ y in the central
%region is given by
%\begin{equation}
%\Delta S = \pi R^2 s_0\tau_0\Delta y
%\end{equation}
%On  the  other  hand  the entropy is in the final state carried
%mostly by pions which are nearly massless bosons; their entropy
%is thus related to their number by
%$\Delta S \approx 3.6\Delta N $. We have therefore
%\begin{eqnarray}
%s_0\tau_0  &=&  {2\pi^4\over  45\zeta(3)}{dN\over  dy}{1\over \pi
%R^2}\nonumber\\
%&\approx &{3.6\over \pi R^2}{\Delta N\over\Delta y}
%\end{eqnarray}
%If the system is initially in aquark-gluon plasma state we have 
%\begin{equation}
%T_i=\left[{2\pi^4\over  45\zeta(3)} {dN\over dy}{1\over 4a_q\pi
%R^2 \tau_0}\right]^{1/3}
%\end{equation}
%The temperature as a function of time is shown in figure 8.
%%
%For $Pb-Pb$ we have taken $dN_{ch}/dy = 550$.
%The  freeze-out  surface  is  shown in figures 9 and 10 for the
%four different scenarios.
%%
%%
\section{Results}
The  transverse  momentum  distribution  of $\pi^+$ is shown in
figure  11a  for  the case where a phase transition occurs at a
temperature of $T_c = $ 160 MeV. In figure 11b we show the same
in  the  absence  of   phase transition. By comparing the two
figures  it  is  clear  that  the absence of a phase transition
leads to much higher transverse momenta.
In  figures 12 we show the distribution of $\pi^+$ for the case
where  the  hadronic  gas  contains  all  known resonances.
The  main  difference  between  the  scenarios  is that the gas
containing  the  full set of resonances is much more sensitive
to  the freeze-out temperature. 
%In the full resonance gas, many
%of the final state pions are due to decays of heavy resonances.
%In  this  case  the  final  momentum  of the resulting pions is
%always smaller than the momentum of the parent resonances. Thus
%the resulting lower transverse momentum in figures 12a and 12b
%while  11a  and 11b both have higher transverse momenta. As the
%freeze-out  temperature  is  lowered  this  effect  is  further
%enhanced.  This explains the differences between figures 11 and
%12.

Figures 13 and 14 show the distributions in transverse mass
for  the  $K^+$.  The  highest  transverse  momenta are seen in
figure  13b  where  we  have 
no slowing down of the transverse expansion due
to  the  phase  transition. Figures 14a and 14b both show lower
transverse  momenta than figures 13a and 13b but they also show
a  greater  sensitivity  to the freeze-out temperature. This is
again  due  to the presence of large numbers of kaons resulting
from decays of heavy resonances.

Figure  15  summarizes  the  average transverse momenta for the
full resonance gas. In figure 15a we show this for the scenario
where  no  phase  transition  occurs, figure 15b shows the case
where  we do have a transition. The phase transition results in
much lower transverse momenta for the outgoing particles.
\section{Thermal Photon Yield}
In the previous section we have analyzed the influence and the role of
the equation of state on the  properties and  behavior
of particle spectra. Particle production, however, is not the only
observable which could give    some insight into the nature
and the structure of the hot hadronic matter which is produced
in ultra-relativistic heavy ion collisions.
In particular photons are a very promising probe in the experimental
search for the QGP formation in A-A collisions 
\cite{ruuskanen,kampf,tr,dks1,dks2,dks3,dks4,dks5}. This is mainly
because the mean free path of a photon in a thermal medium
is expected to be larger than the size of the medium, at least
when the energy of the $\gamma$ is not too small. Thus, photons,
once produced inside the medium will carry direct information
about its nature and its properties. Photons are radiated both
from the plasma and from the hadronic phase. Thus, to find the possible
kinematical window where  photon production from QGP
overwhelms the  hadronic  contribution, one needs to study
different contributing reactions to the overall thermal rate.
Clearly, the thermal photon spectra  are
influenced by the  equation of state of hot hadronic matter as well.
In this section we will discuss the
extent to which the
equation of state could  modify the  thermal photon yield.
We will restrict our discussion  to hard photons with 
energies larger that the temperature of the medium

The Born calculation of the hard photon rate with a bare internal
quark propagator for the $\bar q q\to g\gamma$ and $qg\to q\gamma$
processes leads to divergent production rate in the limit
of vanishing quark mass \cite{1a}. It has been shown,
 however, \cite{baier,2a} that
in the framework of the re-summed perturbative expansion of Braaten and
Pisarski \cite{3a} the mass singularities due to the exchange of massless
quarks are shielded by  Landau damping effects. In a QGP of
temperature $T$, the thermal photon rate has been found t
o be~\cite{baier,2a}:
\begin{equation}
E{{dN^{\gamma}}\over d^4x d^3q}\simeq
 { {e_q^2
\alpha\alpha_s }
\over {2\pi^2}}
T^2
\exp {(-E/T)}   ~
\ln {({{0.23E }\over {\alpha_s T}})}   ~,
\end{equation}
with  $\alpha_s(T)$ being  the strong coupling constant  which
following ref.~\cite{4a}  we  have   parameterized as:
\begin{equation}
\alpha_s(T) =
{{6\pi }\over {(33-2n_f)\ln(8T/T_c)}}
\end{equation}
where the critical temperature is taken to be $T_c = 160$ MeV.
We used  $n_f=2$ for the simplified equation of state
and $n_f = 3$ in the case of all hadrons.
\par
The above  result for the thermal photon rate in a plasma is only
valid for hard $E>T$ photons, and cannot be applied to the soft part of
the spectrum. For soft photons ($E\sim g T$, where $g$ is the QCD
coupling constant ),  one needs to re-sum not only the propagators
but also the dressed vertices. The detailed calculation of the soft photon
rate, performed in \cite{5a}, show that the Braaten-Pisarski re-summation
does not lead to a finite result, free from mass singularities.
Moreover, in \cite{6a}
it was indicated that,   the Braaten-Pisarski re-summed
contribution is         subleading, as there
is a set of singular bremsstrahlung type diagrams which dominate
the soft photon rate.

Having in mind the above problems we will analyze only the
production of hard photons. 
The contribution from quark bremsstrahlung~\cite{7a} is also
neglected in our discussion.

In the hadronic gas  at temperature $T$ the thermal photon
rates were analyzed   in \cite{7a} where all  contributions
originating from the  two loop approximation of the real photon
self energy   using  model with $\pi -\rho  $ interactions
were calculated.
The dominant contribution to the photon rate was  found to be
Compton $\pi\rho\to\pi\gamma$, decay
$\rho\to\pi\pi\gamma$ and the
annihilation $\pi\pi\to \rho\gamma$,
 processes. Moreover, for high enough
temperature, the scattering and the decay  of the  $A_1$
axial vector meson  has been shown to play an essential role
when discussing the thermal photon rate in a hadronic medium.
All these contributing effects are included in our
analysis of thermal photons originating from hadronic matter.
\\
Having established the production cross sections, the photon
$p_T$-spectrum is then obtained  by convoluting
the photon rates with the space-time history of the system as
\begin{equation}
{{dN}\over {d^2p_T dy}}=
\int \tau d\tau rdr d\phi d\eta
\left[ f_Qq_0{{dN^q}\over {d^4xd^3q}} +
 (1-  f_Q)q_0{{dN^\pi}\over {d^4xd^3q}}\right]
 \end{equation}
 where the function $f_Q(r,\tau ,\eta ,\phi)$ describes
  the fraction of the quark
 gluon plasma in the system~\cite{1a}.  In the scenario
  without phase transition
$f_Q\equiv 0$.

The numerical integration  in the above equation is based on the
hydrodynamical scheme summarized in our previous sections
and described in more detail in ref.~\cite{ruuskanen,tr2}.

The results on photon yields in thermal expanding medium
are shown      in figures 16-17.
In figure  16 the  photon $p_T$ spectra were calculated assuming
the hadronic gas  as composed of only
few hadrons and resonances $\pi ,\rho ,\omega ,\eta$.
As it can be seen in fig.16 the photon yield in the scenario
without phase transition exceeds substantially the result
obtained under the assumption of a QGP formation in the initial state.
This dramatic difference arises mostly from the very different
initial temperature required in both  scenarios to reproduce the same final state
particle multiplicity. As it can be deduced from figure 8 one needs
the initial temperature $T_0\sim 318$ MeV in hadronic medium and
$T_0\sim 210$ MeV
in a QGP to get $dN_{ch}/dy \sim 550$.
As the thermal production is a strongly increasing function
of $T$, larger temperatures naturally lead to a higher photon yield.
%
%The following lines were taken out by Dinesh
%
%Also the lifetime of the system is much longer in the scenario
%without phase transition. Consequently both the higher initial temperature and
%the longer lifetime are the responsible parameters leading
%to large differences in the behaviour of the photon yield with and
%without phase transition.
%
%The above lines were taken out by Dinesh
%
%
We have to note, however, that at high enough $T$ the approximation of the
hadronic equation of state as
being  composed only from very few hadrons  is not realistic.
When increasing the temperature beyond 100 MeV the contribution
of   heavier particles and resonances has to be included.
This can  already be seen
in figure 4 where at $T\sim$ 150 MeV the contribution of  resonances increases
 the energy density  of a thermal medium by a factor two.
 Thus it is clear that in modeling the equation of
 state of hadronic matter we should not neglect heavier resonances.

In figure 17 we have made a similar calculation to the one presented 
in figure 16
but this time including all resonances in the hadronic equation of state.
One can see in figure 17 that this time, contrary to figure 16, 
 there is no difference
between the scenario with and without phase transition. In both
cases the hydrodynamical model leads to  almost the same
result  for the photon $p_T$ distribution  in a wide  $p_T$ range.
This is because, contrary to figure 16, the initial temperature and 
the lifetime of the  systems are quite similar. The only difference
between both scenarios is thus the photon production  cross section.
It was shown, however,  \cite{2a} that at   fixed temperature the photon emissivity
of the  QGP plasma and  hadronic matter is quantitatively very similar. That is
why also in the expanding medium the differences between both scenarios
are rather modest as seen in figure 17. A similar conclusion was reached recently
 in reference~\cite{sollfrank}.

Discussing photon production in the
resonance gas we have only included photons originating from the
processes involving pions, rho and $A_1$ vector mesons.
Thus, the results in fig.17 should be considered as the  lower limit for  thermal
photon production. The contribution of  other hadrons,          present
in the hadronic phase, will certainly increase the photon rate both
in the scenario with and without phase transition. We do  expect, however,
that the relative difference between  these scenarios will not be much
affected by the  additional hadronic processes for direct photon production.
\\
%%%%%
%
%
% Added This should be stated very clearly  Dinesh%%%%%%%%%%%%%%%%

 We should point out that the scenario which does not involve a
phase transition has very high particle density in the initial
stages. One could attempt to correct for this by including the
finite size corrections using one of the many Ans\"atze in the literature
for this purpose. We have verified that none of the Ans\"atze 
remain valid at $\tau_i$= 1 fm/$c$, when the volume is only
about 160$\times \Delta y$ fm$^3$. Neglecting the finite volume,
this means more than 10 hadrons/fm$^3$ in the case of the hadronic
gas with only $\pi$, $\rho$, $\omega$, and $\eta$ mesons, and 
more than 3 hadrons/fm$^3$ in the case of exhaustive list of
resonances (see figure 6). This criticism also applies to
the case of particle spectra as well.
%
%
%
%%%%% the above should be stated very clearly  Dinesh%%
%%%
%%%%%%%%%%%

%Analyzing photon emission   we  concentrate entirely on the
%thermal photon rate. This is  because our main scope was to study  the
%dependence of thermal production on the
%    equation of state of strongly interacting matter. 
As a closing remark we
 have to point out however that there is a huge
non-thermal background due to hadron decays,
 mainly $\pi^0$ and $\eta$,
which give substantial contributions to the overall measured photon
distributions. In order to disentangle the quark-gluon plasma presence,
one needs of course first to see thermal production to exceed
the  non-thermal
background.
Recent  comparisons of thermal hydrodynamical 
models~\cite{dks5,8a} with experimentally
measured photon yield by
WA80 collaboration in S-Au collisions at SPS energy has shown
a large contribution of thermal photons. Thus, there are
very  good perspectives to disentangle the QGP in A-A collisions
measuring  photon spectra.
\section{$\eta/\pi$ Ratio}
All measurements of single photons  in a given $p_T$ window
necessarily entail a precise
determination of  the decay photons from $\pi^0\to \gamma \gamma$ 
and $\eta \to \gamma \gamma$ in the same window.
This in turn provides the spectra
of $\pi^0$ and $\eta$ particles, which are often used to
provide a confirmation of the so-called ``$m_T$-scaling''. 
The $m_T$-scaling plays an important role in these studies
by providing a basis for extrapolation of the $p_T$ spectra
of the $\eta$ into unmeasured regions and also to estimate
the $p_T$ spectra for other unmeasured hadrons which give
small photon decay contributions.

In figures 18 and 19 we show the $\eta/\pi^0$ ratio as a function
of  transverse  momentum for different values of the freeze-out
temperature.  Again we compare the cases with and without phase
transition  and  also  the  resonance-rich versus the resonance
poor  hadronic  gas.  We note that in the case where the gas
contains only a few resonances, this ratio rises more rapidly
when there is a phase transition. The  results are essentially
identical for the phase transition and the no phase transition case
when we have the full list of hadrons describing the hadronic
matter.
\section{Conclusions}
We have analyzed  particle and  photon production in 
strongly interacting matter as produced in 
ultra-relativistic heavy ion collisions. Using  
models with transverse flow and a longitudinally 
boost invariant expansion in the z-direction we have studied the 
dependence of the results on the equation of state. For the hadronic 
matter we used two different 
approximations to describe the thermodynamical 
properties of the medium. The model 
with  $\pi, \rho, \omega$, and $\eta$ 
mesons ({\it simplified equation of state}) and the model with all 
hadrons and resonances from the particle data table. Both these 
equations of state have been considered with, and without, quark gluon 
plasma formation in the initial state. In all cases the initial 
temperature was fixed by requiring $dN_{ch}/dy \sim$ 550 in 
the  final  state  and assuming entropy conservation during the
expansion.
The initial thermalization time $\tau_{0} \sim$ 1 $fm$ was also 
assumed.

It was shown that most physical observables derived within the above 
hydrodynamical model for the expansion dynamics are very sensitive to 
the equation of state, and assumptions were made about the nature of 
the initially created fire cylinder. This is particularly evident 
when comparing the results of the simplified equation of state in the 
scenario with, and without, phase transition. Here the hadronic gas 
scenario leads to a substantially higher rate for 
the $p_{T}$ distribution of both particles and photons. 
This is entirely due to 
a very different initial temperature required in both of these cases. 
%
% Modified 
%
%
%In 
%any case the simplified hadronic equation of state with only a few 
%particles is highly non-realistic and should not be applied when 
%discussing any thermal production in ultra-relativistic heavy ion 
%collisions as it leads to misleading conclusions. This is 
%particularly true for the scenario without phase transition.
%
% Modification follows:
%
%
In  any  case  the  simplified hadronic equation of state with
only   a  few  particles  becomes  unrealistic  at  higher
temperatures  and can lead to misleading conclusions if applied
at  the high  temperatures,  implicit  in  the no phase
transition scenario.
%
% End of modification
%
%

When assuming more complete equation of state with a few hundred 
hadrons and resonances the difference between the scenario with, and 
without, phase transition is rather modest. Both photon and the
particle spectra in the wide $p_{T}$ range, show quantitatively very similar 
behaviour. Some small excess of the yield in the scenario without 
phase transition reflects slightly higher initial temperature than 
required for the quark gluon plasma formation in the initial state. 
Thus, from the above, one could conclude that from the measured 
$p_{T}$ spectra it would be rather hard to disentangle the quark 
gluon plasma formation in the initial state. We have to repeat, 
however, that there are  conceptual difficulties in applying only 
hadron gas equation of state for the conditions of the SPS-energy. At 
the temperature $T \sim$ 200 MeV, required to satisfy measured final 
state pion multiplicity in Pb-Pb collisions, we are ending up with 
a particle density of the order of 2--3/fm$^3$. When considering 
hadrons as extended objects of the radius $R \sim$ 1 fm, these initial
conditions are not acceptable. This is in the spirit of the Hagedorn
model which leads to the appearance of the limiting temperature above 
of which the hadron concept does not make any sense. Thus, from the 
 point of view of principle, for the SPS energy the assumption of the 
equation of state with the phase transition scenario and the quark 
gluon plasma formation in the initial state seems to be more 
natural.

Discussing particle spectra for the conditions of Pb-Pb collisions at 
the SPS energy we have not made any comparisons with the existing 
experimental data of NA49 collaboration. The main reason for this was 
due to the simplification of our calculations which restrict us only to 
direct thermal particle production without taking into account the 
decay of resonances. This approximation is sufficient to study the 
role of the equation of state but is not adequate to make any 
comparisons with the experimental data.
%
%
% New addition
%
%
It is quite clear that as long as the temperatures remain small
compared  to  the pion mass, one may use either of the hadronic
equations  of  state as the contributions of the heavier mesons
will  be negligible. As the temperature rises, the contribution
of the heavier resonances will increase. If we limit the number
of  allowed  particles  to a small number, as in the simplified
equation  of  state,  the  temperature  becomes very high for a
given entropy density. On the other hand, the equation of state
having  the  full hadronic spectrum implies a complete chemical
equilibrium  amongst  all  hadrons,  which  may  or  may not be
achieved. We have already stated repeatedly that both equations
of  state imply too high a density in the initial stage for the
case  without  phase transition. It is of considerable interest
to  note  that if we assume the initially produced matter to be
in  the QGP phase, then the results for the thermal photons and
also  for the other quantities that we have studied differ only
marginally for the different equations of state considered.

In summary, we have shown that most observables are very sensitive to 
the equation of state.
For the simplified equation of state  the hadronic gas 
scenario (without phase transition) 
leads to the  highest rate for the 
$p_T$-distribution
of all particles. 
Using  the  equation  of state with all particles listed in the
particle data booklet, the difference 
found between the scenarios with and
without phase transition is very modest. Both photon and 
particle spectra, in a wide $p_T$ range, show very similar 
behavior. 
We therefore conclude that 
if  we  can reconcile with very large particle densities in the
hadronic gas description, then from the 
$p_T$ spectra alone, it will be hard to 
prove  the presence of a quark 
gluon plasma.\\[.5 cm] 
%
%
%
%
%
%
%
%%Do you want to add this %%%%%%%%%%%%%%%%%%%%%%%%
%
% It is often assumed that this description could still be used to
% evaluate the particle spectra at larger $p_T$, even with-out
% considering the resonance decay. In view of the large transverse
% flow seen in the present work, and the fact, the heavier
% particles are more strongly affected by the flow, we have attempted
% that either. 
%
%%%%%%%%%Do you want to add this %%%%%%%%%%%%%%%%%%%%%%%%
%
%
%
\acknowledgments
We would like to thank Kevin Haglin for supplying us with
the code used for figure 3. We also thank Vesa Ruuskanen for his help with
solving  the hydrodynamic equations numerically.
One of us acknowledges financial support from the University of Cape Town (URC) and 
the Foundation for Research Development (FRD, Pretoria).
Two  of  the  authors (K.R. and D.K.S.) would like to thank the warm
and  generous  hospitality of the University of Cape Town where
this  project  was  conceived  and  largely executed. K.R. also
acknowledges   partial   support   of  the  Gesellschaft  f\"ur
Schwerionenforschung   (GSI)  and  the  Committee  of  Research
Development (KBN 2-P03B-0998).
\begin{center}
FIG. 1. The average transverse momentum as a function of the
freeze-out temperature in the absence of transverse flow.
\end{center} 
\begin{center}
FIG. 2a. The average transverse momentum as a function of the
freeze-out  temperature  in  the  presence of thermal flow. The
hadronic  gas  is  made  up  of  $\pi, \rho, \omega$ and $\eta$
mesons and no phase transition takes place.
\end{center}
\begin{center}
FIG. 2b. The average transverse momentum as a function of the
freeze-out  temperature  in  the  presence of thermal flow. The
hadronic  gas  is  made  up  of  $\pi, \rho, \omega$ and $\eta$
mesons and a phase transition takes place at $T_c$ = 160 MeV.
\end{center}
\begin{center}
FIG. 3. The mean free path of pions and kaons as a function of
temperature (from reference [Kevin Haglin]). 
\end{center}
\begin{center}
FIG. 4. The energy density as a function of the temperature.
The  dashed  line  corresponds  to  a gas containing only $\pi,
\rho,  \omega$  and  $\eta$  mesons  (with  and  without phase
transition  to  a quark gluon plasma). The full line takes into
account  all  hadronic  resonances  with  masses  below 2.5 GeV
(again with and without phase transition).
\end{center}
\begin{center}
FIG. 5. The   speed  of  sound  (squared)  as  a  function  of
temperature  in a hadronic gas (without phase transition). 
\end{center}
\begin{center}
FIG. 6. The  particle density as a function of temperature for
two  different  scenarios  expalined  in the text. No
repulsive  effects  or  excluded  volume  corrections have been
taken into account.
\end{center}
\begin{center}
FIG. 7. Pressure   as   a  function  of  temperature  for  two
different hadronic scenarios. No
repulsive  effects  or  excluded  volume  corrections have been
taken into account.
\end{center}
\begin{center}
FIG. 8. The  temperature  as a function of the proper time for
the four different scenarios considered in the text. 
\end{center}
\begin{center}
FIG. 9a. The freeze-out time as a function of the radial
distance $r$ for
different values of the freeze-out temperature $T_f$.
The  hadronic  gas contains only $\pi, \rho, \omega$ and $\eta$
mesons and no phase transition takes place. 
\end{center}
\begin{center}
FIG. 9b. The freeze-out time as a function of the radial
distance $r$ for
different  values  of  the freeze-out temperature $T_f$. The
phase transition occurs to a quark-gluon plasma occurs at $T_c$
=  160 MeV. 
\end{center}
\begin{center}
FIG. 10a. Freeze-out  time  as  a  function  of the radial
distance for a 
resonance gas without phase transition to a quark-gluon plasma.
 Different values of the freeze-out
temperature are indicated. 
\end{center}
\begin{center}
FIG. 10b. Freeze-out  time  as a function of the radial distance
$r$ for a 
resonance  gas  with  a  phase transition occuring at 
$T_c$ = 160 MeV. Different values of the freeze-out
temperature are indicated.
\end{center}
\begin{center}
FIG. 11a. Distribution in transverse mass at y=0 of pions for a gas
containing   only   $\pi,  \rho,  \omega$  and  $\eta$  mesons.
Different  values  of the freeze-out temperature are indicated.
Phase transition occurs at $T_c = 160$ MeV.
\end{center}
\begin{center}
FIG. 11b. Distribution in transverse mass at y=0 of pions for a gas
containing   only   $\pi,  \rho,  \omega$  and  $\eta$  mesons.
Different values of the freeze-out temperature are indicated.
\end{center}
\begin{center}
FIG. 12a. Distribution in transverse mass at y=0 of pions for a gas
containing   all known resonances with masses below 2.5 GeV.
Different values of the freeze-out temperature are indicated.
The phase transition occurs at $T_c = 160$ MeV. 
\end{center}
\begin{center}
FIG. 12b. Distribution in transverse mass at y=0 of pions for a gas
containing   all known resonances with masses below 2.5 GeV.
Different values of the freeze-out temperature are indicated.
\end{center}
\begin{center}
FIG. 13a. Distribution in transverse mass at y=0 of
 kaons for a gas
containing   only   $\pi,  \rho,  \omega$  and  $\eta$  mesons.
Different  values  of the freeze-out temperature are indicated.
Phase transition occurs at $T_c = 160$ MeV. 
\end{center}
\begin{center}
FIG. 13b. Distribution in transverse mass at y=0 of kaons for a gas
containing   only   $\pi,  \rho,  \omega$  and  $\eta$  mesons.
Different values of the freeze-out temperature are indicated.
\end{center}
\begin{center}
FIG. 14a. Distribution in transverse mass at y=0 of kaons for a gas
containing   all known resonances with masses below 2.5 GeV.
Different values of the freeze-out temperature are indicated.
Phase transition occurs at $T_c = 160$ MeV. 
\end{center}
\begin{center}
FIG. 14b. Distribution in transverse mass at y=0 of kaons for a gas
containing   all known resonances with masses below 2.5 GeV.
Different values of the freeze-out temperature are indicated.
\end{center}
\begin{center}
FIG. 15a. Average   transverse   momentum   
of $\pi^+$, $K^+$ and nucleons as   a  function  of
the freeze-out temperature $T_f$ for a gas containing 
all  resonances  with masses up to 2.5 GeV. No phase transition
takes place.
\end{center}
\begin{center}
FIG. 15b. Average   transverse   momentum 
of $\pi^+$, $K^+$ and nucleons as   a  function  of
the freeze-out temperature $T_f$ for a gas containing 
all  resonances  with  masses up to 2.5 GeV. A phase transition
has been built in at $T_c=160$ MeV. 
\end{center}
\begin{center}
FIG. 16. Distribution  in  transverse  momentum at y=0 of
photons for a gas
containing   only $\pi, \rho, \omega$ and $\eta$ mesons
with (dashed line) and without (full line) phase transition.
\end{center}
\begin{center}
FIG. 17. Distribution  in  transverse  momentum at y=0 of
photons for a gas
containing   all known resonances with masses below 2.5 GeV
with (dashed line) and without (full line) phase transition.
\end{center}
\begin{center}
FIG. 18a. The  $\eta/\pi$  ratio  as  a  function  of the
transverse momentum $p_T$  for a gas
containing only $\pi, \rho, \omega$ and $\eta$ mesons.
Different values of the freeze-out temperature are indicated.
\end{center}
\begin{center}
FIG. 18b. The  $\eta/\pi$  ratio  as  a  function  of the
transverse momentum $p_T$ for a gas
containing only $\pi, \rho, \omega$ and $\eta$ mesons.
Different values of the freeze-out temperature are indicated.
\end{center}
\begin{center}
FIG. 19a. The  $\eta/\pi$  ratio  as  a  function  of the
transverse momentum $p_T$  for a gas
containing   all known resonances with masses below 2.5 GeV.
Different values of the freeze-out temperature are indicated.
\end{center}
\begin{center}
FIG. 19b. The  $\eta/\pi$  ratio  as  a  function  of the
transverse momentum $p_T$  a gas
containing   all known resonances with masses below 2.5 GeV.
Different values of the freeze-out temperature are indicated.
\end{center}
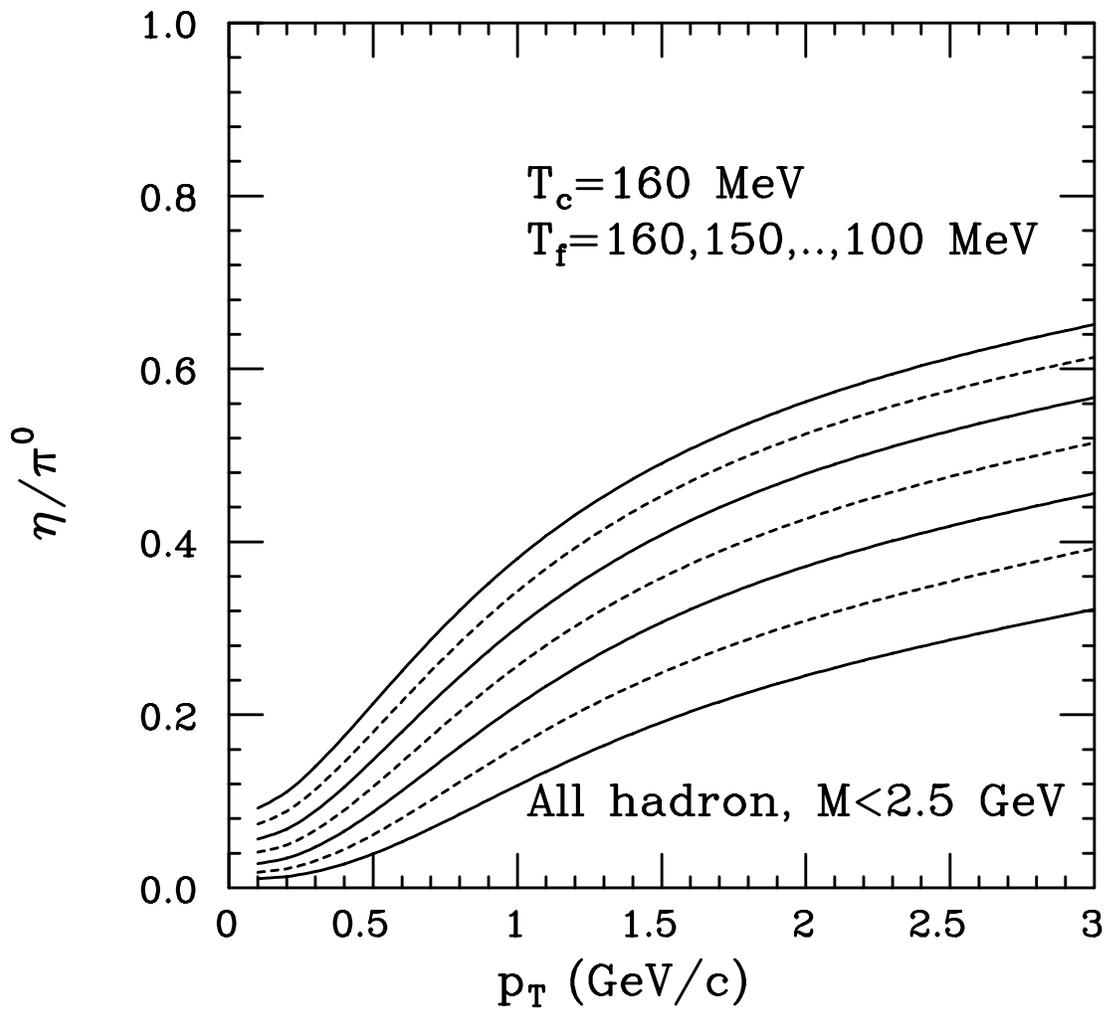
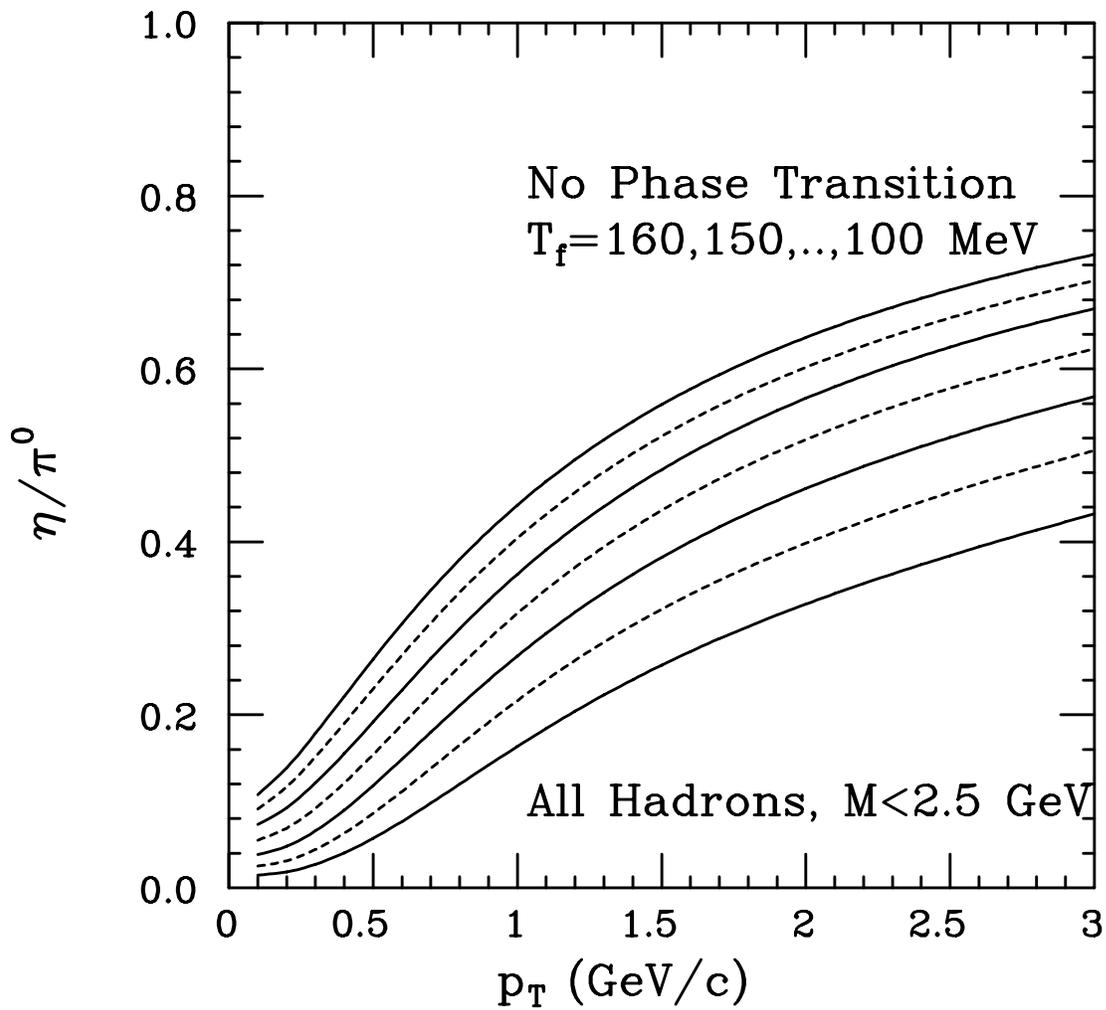
\end{document}